\documentclass[12pt,preprint]{aastex}

\shorttitle{slipping reconnection as a precursor to flare}
\shortauthors{Li et al.}

\begin{document}

\title{Slipping Magnetic Reconnection of Flux Rope Structures as a Precursor to an Eruptive X-class Solar Flare}

\author{Ting Li\altaffilmark{1,2}, Kai Yang\altaffilmark{3,4}, Yijun Hou\altaffilmark{1,2} \& Jun Zhang\altaffilmark{1,2}}

\altaffiltext{1}{Key Laboratory of Solar Activity, National
Astronomical Observatories, Chinese Academy of Sciences, Beijing
100012, China; liting@nao.cas.cn} \altaffiltext{2}{University of
Chinese Academy of Sciences, Beijing 100049, China}
\altaffiltext{3}{School of Astronomy and Space Science, Nanjing
University, Nanjing 210023, China} \altaffiltext{4}{Key Laboratory
for Modern Astronomy and Astrophysics (Nanjing University), Ministry
of Education, Nanjing 210023, China}

\begin{abstract}

We present the quasi-periodic slipping motion of flux rope
structures prior to the onset of an eruptive X-class flare on 2015
March 11, obtained by the \emph{Interface Region Imaging
Spectrograph} (\emph{IRIS}) and the \emph{Solar Dynamics
Observatory} (\emph{SDO}). The slipping motion occurred at the north
part of the flux rope and seemed to successively peel off the flux
rope. The speed of the slippage was 30$-$40 km s$^{-1}$, with an
average period of 130$\pm$30 s. The Si {\sc iv} 1402.77 {\AA} line
showed a redshift of 10$-$30 km s$^{-1}$ and a line width of
50$-$120 km s$^{-1}$ at the west legs of slipping structures,
indicative of reconnection downflow. The slipping motion lasted
about 40 min and the flux rope started to rise up slowly at the late
stage of the slippage. Then an X2.1 flare was initiated and the flux
rope was impulsively accelerated. One of the flare ribbons swept
across a negative-polarity sunspot and the penumbral segments of the
sunspot decayed rapidly after the flare. We studied the magnetic
topology at the flaring region and the results showed the existence
of a twisted flux rope, together with quasi-separatrix layers (QSLs)
structures binding the flux rope. Our observations imply that
quasi-periodic slipping magnetic reconnection occurs along the
flux-rope-related QSLs in the preflare stage, which drives the later
eruption of the flux rope and the associated flare.

\end{abstract}

\keywords{magnetic reconnection --- Sun: flares --- Sun: UV
radiation--- Sun: transition region}

\section{Introduction}

Solar flares, often associated with filament eruptions and coronal
mass ejections (CMEs), are major drivers of space weather (Gosling
et al. 1991). During these events, magnetic free energy is converted
to radiation, energetic particle acceleration and kinetic energy of
plasma through magnetic reconnection (Forbes et al. 2006). The
two-dimensional (2D) standard flare model (Shibata \& Magara 2011)
was applied to explain many flare phenomena. In this model, magnetic
reconnection takes place at the magnetic null-point under the
eruptive flux rope and flare ribbons, cusp-shaped loops and
post-flare loops are thus formed (Schmieder et al. 1996; Milligan \&
Dennis 2009). Nevertheless, actual flares are intrinsically
three-dimensional (3D) events (Janvier et al. 2015). The 2D standard
flare model may not explain some features of flares (Wang \& Liu
2012), such as the morphology of the ribbons and the motions of
small-scale bright knots along ribbons. Therefore, a comprehensive
understanding of the 3D physical processes of flares is important,
particularly magnetic reconnection.

By analyzing the magnetic topology and geometry of the flaring
region, the locations where magnetic reconnection could occur can be
well understood. In 2D, magnetic reconnection is initially
considered to occur at a null point (Sweet 1958), where the magnetic
field vanishes. A separatrix surface is another topological feature
preferential for forming the strong electric current sheets, through
which the field line connectivity is discontinuous (Longcope 2005).
Gorbachev \& Somov (1988) first found the correspondence of the
separatrices obtained from the potential magnetic field model and
flare ribbons. Afterwards, several studies found the relationship
between the observed flare ribbons and the computed separatrices
from potential and linear force-free field models (Demoulin et al.
1993; van Driel-Gesztelyi et al. 1994; Mandrini et al. 1995, 2014).
However, the flares without null points and the associated
separatrices are observed (Demoulin et al. 1994). Therefore, Priest
\& D{\'e}moulin (1995) and Demoulin et al. (1996) introduced the
concept of quasi-separatrix layers (QSLs), where the magnetic
connectivity shows strong gradients, but is still continuous. The
magnetic field distortion is defined by a strong value of the
squashing degree Q (Titov et al. 2002; Titov 2007). QSLs are
preferential locations for magnetic reconnection because high
electric current density regions can be formed at QSLs during the
magnetic field evolution (Aulanier et al. 2005; Wilmot-Smith et al.
2009). Savcheva et al. (2012, 2015) found the close correspondence
between flare ribbons and the locations of QSLs obtained from the
data-constrained nonlinear force-free field (NLFFF) models created
with the flux rope insertion method.

When magnetic reconnection occurs in the QSLs, the field lines
crossing the QSLs exchange their connectivity with the neighboring
fields and their motion is seen as an apparent slipping motion of
field lines (Priest \& D{\'e}moulin 1995; Demoulin et al. 1996;
Aulanier et al. 2006, 2007; Masson et al. 2009). Priest \&
D{\'e}moulin (1995) analytically predicted that the magnetic
reconnection in the QSLs is characterized by the flipping of
magnetic field lines as they slip rapidly through the plasma. If the
speed of the apparent motion of the field lines is at
super-Alfv\'{e}nic time scales, the process is the so-called
``slip-running reconnection." While if the change of connectivity is
sub-Alfv\'{e}nic, this is said to be ``slipping reconnection"
(Aulanier et al. 2006). In the 3D magnetohydrodynamic (MHD)
simulation of Janvier et al. (2013), the slipping motion speed is
not constant as time goes by and it could be sub- or
super-Alfv\'{e}nic. The simulation of Janvier et al. (2013)
generally matches the recent observations of Li \& Zhang (2014), who
presented the slippage of flux rope structures along a hook-shaped
flare ribbon. The slipping motion delineated a ``triangle-shaped
flag surface", implying one-half of the QSL structure. Observational
studies showed the slipping motion of individual flare loops and
small-scale bright knots in flare ribbons during eruptive flares
(Dud{\'{\i}}k et al. 2014, 2016; Zheng et al. 2016; Sobotka et al.
2016), which satisfies the slipping reconnection regime. Based on
the observations from the \emph{Solar Dynamics Observatory}
(\emph{SDO}; Pesnell et al. 2012) and the \emph{Interface Region
Imaging Spectrograph} (\emph{IRIS}; De Pontieu et al. 2014), Li \&
Zhang (2015) found that the flare loops and flare ribbon
substructures both exhibited the quasi-periodic slipping motion with
a period of about 3-6 min.

The direct observations about the slipping nature of 3D magnetic
reconnection are very rare due to the low spatial resolution and
limited channels of previous instruments. Recently, the rich
high-quality observations provide us a chance to study the 3D
evolution process of flares. In this paper, we report that the flux
rope structures exhibited the quasi-periodic slippage along the
flux-rope-related QSLs in the preflare stage, which drives the
initiation of an X2.1 flare on 2015 March 11 and the later eruption
of the flux rope. The previous studies reported the slipping motions
of flare loops and flux rope structures during the flare process,
however, the preflare slipping motion of flux rope structures has
never been reported before. The outline of the paper is as follows:
the observations and data analysis are presented in Section 2,
Section 3 presents the results of preflare activities and the later
eruption, Section 4 shows the summary and discussion.

\section{Observations and Data Analysis}

We combine data from the \emph{SDO} and the \emph{IRIS} to
investigate the preflare slipping motion of flux rope structures and
the later eruption process. Full sun images from the Atmospheric
Imaging Assembly (AIA; Lemen et al. 2012) are available with a
resolution of $\sim$0$\arcsec$.6 per pixel and a cadence of 12
seconds for the EUV passbands. The observations of AIA 1600 {\AA},
304 {\AA} and 131 {\AA} passbands are used. The full-disk
line-of-sight (LOS) magnetograms and the photospheric vector
magnetic field data of the AR observed by the Helioseismic and
Magnetic Imager (HMI; Scherrer et al. 2012) are also applied. The
\emph{IRIS} 1330 {\AA}, 1400 {\AA} and 2832 {\AA} slit-jaw images
(SJIs) cover the majority of the active region (AR), with a spatial
sampling of $\sim$0$\arcsec$.33 per pixel and a cadence of about 20
seconds for each passband. The spectral data are taken in a 4-step
raster mode with steps of 2$\arcsec$, giving a total field of view
(FOV) of 6$\arcsec$$\times$ 119$\arcsec$. Each raster step takes
about 5 s (2 s exposure) and the spectral sampling is $\sim$0.025
{\AA} pixel$^{-1}$. We analyze the spectroscopic observations of the
Si {\sc iv} 1402.77 {\AA} line formed in the transition region with
a temperature of $\sim$ 80000 K (Tian et al. 2014) and apply a
single-Gaussian fit to obtain the Doppler shift and line width at
the locations of the slipping structures (Peter et al. 2014).

\section{Results}

\subsection{Overview of Productive AR 12297}

The target flare of this study occurred in NOAA AR 12297 in the
southern hemisphere on 2015 March 11. This AR produced many flares
including one X- and 18 M-class events since its appearance at the
east limb on March 07. Before the occurrence of the X2.1 flare, two
M-class flares (M2.9 and M2.6 in Figures 1(a)-(b)) occurred at
around 00:00 and 08:00 UT on March 11. The region also produced
several C-Class flares during 5 hours before the X-class flare. The
small flares may reduce the constraint of the flux rope system by
the rearrangement of magnetic fields, which makes it easier for the
later eruption of the flux rope and the X-class flare. A bipolar
pair (P1$-$N1) emerged in the AR center and the emerging
opposite-polarity magnetic flux gradually separated from each other
(Figures 1(d)-(f)). Simultaneously, strong shearing motions were
observed with positive patch P1 continuously moving to the northwest
(red arrows) and negative patch N1 drifting to the southeast (green
arrows). A hook-shaped filament (length of $\sim$150 Mm) was located
at the AR (Figure 1(b)), and at the east part of the filament, a
flux rope could be observed in 131 {\AA} channel (Figure 1(c)). The
north end of the flux rope anchored in the main sunspot with
positive-polarity magnetic fields (Figure 1(f)).

\subsection{Slipping Motion of Flux Rope Structures}

At the north part of the flux rope, the 1400 {\AA} and 1330 {\AA}
SJIs showed multiple bright loop structures (Figure 2; see Animation
1400-slippage). These loop structures (dashed curves in Figure 2(d))
were highly sheared around the main sunspot. Starting from about
15:27 UT, the bright structures successively slipped toward the east
and seemed to peel off the flux rope. Two of the slipping processes
from the west to the east were respectively shown in Figures
2(a)-(d) and 2(e)-(h). It is not the same ``clump" of brightness
(pointed by arrows) observed at each time, but new flux rope
structures. The slipping motion was evident at the loop tops and the
eastern footpoints. The eastern footpoints of the loops are
relatively scattered in the east-west direction (Figure 2(f)). While
the western part of the loops is concentrated and the slippage could
not be clearly observed. Note that the term ``slipped" or
``slipping" is only the phenomenological term.

The comparison of HMI magnetograms and 1400 {\AA} images shows that
the west footpoints of the slipping structures are located at
positive-polarity magnetic fields of the main sunspot and the east
footpoints are at negative-polarity magnetic fields (Figures
3(a)-(b)). The slipping structures seem obscure in 304 {\AA} and 131
{\AA} images (Figures 3(c)-(d)), probably because of the low spatial
resolution of AIA EUV data and the blocking effect of filament
materials along the LOS direction. In \emph{IRIS} 1400 {\AA}
observations, the slipping structures appear as bright structures
(Figure 3(a)). However, in AIA EUV images, both the dark and bright
slipping structures are observed (arrows in Figures 3(c)-(d)). We
suggest that the dark structures in 304 {\AA} and 131 {\AA} images
correspond to filament materials, which are not evident in the
optically thin lines such as 1400 {\AA} and 1330 {\AA} formed at
transition region temperatures (Li \& Zhang 2016). At about 15:53
UT, EUV brightenings were observed at the east footpoints of
slipping structures (Figure 3(e)). Then the brightenings extended
towards the southeast and lasted about 4 minutes. At the late stage
of the slipping motion, a fan-shaped surface (delineated by dashed
curves in Figure 3(f)) appeared overlying the filament in the 131
{\AA} channel.

For investigating the kinematic evolution of the slipping
structures, the time-distance plots obtained along the cut``A$-$B"
(Figure 3(a)) are shown in Figure 4. Several moving intensity
features are displayed in the time-distance plots of 1400 {\AA} and
1330 {\AA} (Figures 4(a)-(b)), and each strip denotes the apparent
slipping motion of flux rope structures. The 304 {\AA} time-distance
plot shows alternately bright and dark strips (Figure 4(c)). The
slippage of flux rope structures is almost at a constant velocity of
30$-$40 km s$^{-1}$ and the propagating distance is about 10 Mm. The
appearance of flux rope structures and the slipping motion are
intermittent and quasi-periodic. Between 15:27 UT and 15:48 UT,
about seven bright strips were observed. The 1400 {\AA} and 1330
{\AA} profiles along ``L1" (Figures 4(a)-(c)) are approximately
consistent (Figure 4(d)). The time intervals between two neighboring
peaks range from 102 s to 153 s, and the average period is about 130
s. The small peak just before 15:40 UT in 1400 {\AA} profile was
also selected for calculating the period of the slipping motion. As
shown in Figures 4(a)-(b), there is indeed a weak and thin strip at
this time. There is no clear signature in 1400 {\AA} profile
probably because the smoothing process of horizontal slices along
the ``L1" weakens the signal. Moreover, the peak before 15:30 UT in
1330 {\AA} was not included while analyzing the quasi-periodic
pattern. As seen from the 1330 {\AA} stack plot (Figure 4(b)), this
peak corresponds to the stationary brightening, which is not related
to the slipping structure. The intensity variation in the 304 {\AA}
profile is smaller than 1400 {\AA} and 1330 {\AA} profiles, and
about five peaks could be clearly discerned in the 304 {\AA}
profile.

The spectroscopic properties at the western legs of slipping
structures are investigated and displayed in Figure 5. At 15:37:21
UT, the brightenings at the east end of a loop-like structure were
observed (orange arrow in Figure 5(a)). The intersection of the loop
structure and the \emph{IRIS} slit was at the west leg of the
analyzed structure (orange diamonds in Figures 5(a)-(b)). Applying
the single Gaussian fitting of Si {\sc iv} 1402.77 {\AA} line, we
get a redshift of $\sim$ 18 km s$^{-1}$ at the western leg of the
slipping structure, with the line width of about 50 km s$^{-1}$
(Figure 5(c)). The background location used to correct the center is
not displayed, for it is beyond the FOV of Figure 5. Then the loop
structures continually slipped to the east. At about 15:42:32 UT,
the intersections of two loop structures and the \emph{IRIS} slit
were selected (orange and blue diamonds in Figures 5(d)-(e)).
Meanwhile, the brightenings appeared at the east end of one loop
structure (blue arrow in Figure 5(d)). Similarly, the profiles of
the Si {\sc iv} line at west legs of the two structures are fitted
and both exhibit evident redshifts (Figures 5(e)-(f)). For the north
location (orange diamonds in Figures 5(d)-(e)), the redshift
velocity is about 12 km s$^{-1}$ and the line width is $\sim$ 58 km
s$^{-1}$ (orange curves in Figure 5(f)). The Si {\sc iv} profile at
the west leg of the south brighter structure shows two peaks, with
one peak at the line center and the other at the red wing (blue
solid curve in Figure 5(f)). The line-profile at this location was
fitted by a double-Gaussian function. The first peak comes from the
background emission and the second peak is from the west leg of the
slipping structure. The slipping reconnection at the east part
causes the outflow along the loop structure and thus the line
profile at the west leg is redshifted. The double Gaussian fitting
shows a larger redshift of about 32 km s$^{-1}$ at the south
brighter structure, with the line width reaching about 119 km
s$^{-1}$ (red dotted curve in Figure 5(f)). The uncertainty of the
redshifts is $\sim$ 1 km s$^{-1}$ and that of the widths is about 2
km s$^{-1}$. They are estimated according to the 1-sigma error in
GAUSSFIT function.

\subsection{Magnetic Topology Around the Main Sunspot}

We obtain the 3D coronal magnetic field by the NLFFF extrapolation
method (Wheatland et al. 2000; Wiegelmann 2004). The vector
magnetogram for the extrapolation is at 16:04 UT before the flare
from the Space-weather HMI Active Region Patches, in which the
180\arcdeg\ ambiguity and the projection effect have already been
resolved (Sun et al. 2013; Bobra et al. 2014). Before the
extrapolation, the vector magnetogram data have undergone an
additional preprocessing to remove the net force and torque
(Wiegelmann et al. 2006). Moreover, based on the extrapolated
results, we calculated the 3D squashing factor, $Q$, with the method
proposed by Pariat \& D{\'e}moulin (2012). The $Q$ factor is a
measurement of the gradient of the magnetic field connectivity,
which is usually used to determine the QSL (Titov et al. 2002).

The extrapolation results show the existence of a twisted flux rope
around the main sunspot before the eruption (Figure 6), similar to
the \emph{SDO} and the \emph{IRIS} observations. We only show the
extrapolation results at 16:04 UT prior to the largest flare as we
are focusing on the magnetic topology during the period of preflare
motion. The western footpoints of the flux rope are located at the
positive-polarity sunspot and the eastern ones are at a small
negative-polarity sunspot and facula region. The 3D QSL structures
are surrounding the flux rope. The intersection of the QSLs with the
lower boundary approximately overlaps the locations of flare ribbons
(Figures 6(b) and 7(c)). The correspondence of the intersection of
the QSL with bottom boundary and flare ribbons has been reported
before (Yang et al. 2015; Savcheva et al. 2015; Zhao et al. 2016).
The comparison of the extrapolated magnetic field and the
observations (Figures 2-3) shows that the magnetic field lines with
negative-polarity footpoints along the east-west direction probably
correspond to the slipping structures (white arrows in Figure 6). As
seen in Figures 3(a)-(b), the east ends of slipping structures are
located at the negative-polarity magnetic fields and extend along
the east-west direction. The magnetic field lines at the north of
white arrows in Figure 6 show a similar feature. While the field
lines cross the overlying QSL, they undergo a succession of
reconnection processes and result in the apparent slipping motion
towards the east.

\subsection{Flux Rope Eruption and Associated X2.1 Flare}

The slipping motion of flux rope structures lasted about 40 min from
15:27 UT to 16:07 UT. At the late stage of the slippage, the entire
flux rope became unstable and started to erupt from about 16:00 UT.
Figure 7 shows the multi-wavelength appearance of the eruptive event
and the corresponding magnetogram observed by the \emph{IRIS} and
the \emph{SDO} (see Animations 1400-eruption and 131-eruption). At
the location of the bright flux rope observed in 1400 {\AA} (Figure
7(a)), the dark filament material appeared in the 304 {\AA}
observations and showed a consistent evolution with the flux rope
(Figure 7(b)). The pre-flare EUV brightenings were observed
underlying the flux rope from about 16 min before the flare start
(Figures 3(e) and 7(b)). In order to analyze the kinematic evolution
of the flux rope in detail, we obtain the time-distance plots
(Figure 8) in different wavelengths along the slice ``C$-$D" (Figure
7(d)). The flux rope initially underwent a slow$-$rise phase at a
speed of about 10 km s$^{-1}$. The associated X2.1 flare initiated
at 16:11 UT from the GOES SXR 1$-$8 {\AA} flux (Figure 8(a)). Almost
simultaneously, the flux rope began to impulsively accelerate and
the velocity increased to $\sim$ 170 km s$^{-1}$ at 16:15 UT
(Figures 8(a)-(c)). Extremely bright post-eruption arcades appeared
underlying the erupting flux rope (Figures 7(d)-(e)). The south part
of the flux rope was gradually stretched and the twisted structures
at the south part could only be observed in the channel of 131 {\AA}
compared to \emph{IRIS} 1400 {\AA} and AIA 304 {\AA} (Figures
7(d)-(f)). This implies that the south part of the flux rope has a
very high temperature (the 131 {\AA} channel corresponds to about 11
MK, and also sensitive to plasma about 1 MK as well as above 10 MK;
O'Dwyer et al. 2010), similar to the previous observations of flux
ropes with the \emph{SDO} (Li \& Zhang 2013; Cheng et al. 2014;
Zhang et al. 2015a). At 16:16:30 UT, the top part of the flux rope
underwent an obvious clockwise kink motion (Figures 7(g)-(i)) and
the twist is transformed into the writhe of the axis. The kink
motion implies the occurrence of kink instability (Hood \& Priest
1981; Guo et al. 2010; Yan et al. 2014). The eastern footpoints of
the flux rope are rather extended along the negative-polarity
magnetic fields, while its western footpoints are relatively
concentrated nearby the main positive-polarity sunspot (Figure
7(i)). The associated flare reached its peak at 16:22 UT and ended
at 16:29 UT (Figure 8(a)). The flux rope continued to be accelerated
upward during the impulsive phase of the flare.

Figure 9 shows the evolution of the sunspots at the flaring region
in \emph{IRIS} 2832 {\AA} images. One positive-polarity flare ribbon
(PR; Figures 7(c) and 9(b)) and two negative-polarity ribbons (NR1
and NR2) were involved in the flare. The ribbon PR extended towards
the south at a speed of $\sim$ 8 km s$^{-1}$ and swept across a
fraction of the main sunspot with a distance of about 2.2 Mm. The
ribbons NR1 and NR2 moved in the opposite directions and almost
swept across the entire negative-polarity sunspots. Transient
spike-like brightenings were observed at the east of NR2 while NR2
swept across the north negative-polarity sunspot (Figure 9(c)). Then
the northeast penumbra (area ``A1") started to decay and the
penumbral dark fibrils progressively lost their filamentary
structure (Figures 9(d)-(f)). Finally the dark fibrils completely
disappeared within an hour after the onset of the flare. The size of
the umbra region became smaller just after the flare (green lines in
Figures 9(a) and (d)). Then the area of the umbra gradually
increased during the process of penumbral decay (Figure 9(f)). The
time-distance plot along the slice ``E$-$F" showed the evolution of
penumbral segments after the ribbon NR2 swept across them (Figure
9(g)). Continuous outflow along the penumbral filaments was shown in
the time-distance plot (red dashed lines). This outflow had a
velocity of 1$-$3 km s$^{-1}$, consistent with the typical speed of
the Evershed flow (Evershed 1909). After the flare started, the
penumbra inflow towards the umbra became evident (orange dashed
lines). The inflow had a velocity of about 1 km s$^{-1}$ and existed
in the early stage of penumbral decay. The 2832 {\AA} emission
intensity within the area ``A1" was enhanced by about 40 \% (black
curve in Figure 9(h)). The background region ``A2" away from the
flare brightenings does not show the flaring time profile seen in
``A1" (blue curve in Figure 9(h)).

\section{Summary and Discussion}

We present the \emph{IRIS} and the \emph{SDO} observations of the
slipping motion of flux rope structures in the preflare stage on
2015 March 11. The slippage occurred at the north part of the flux
rope around the main sunspot of AR 12297. Multiple bright loop
structures of the flux rope at 1400 {\AA} and 1330 {\AA}
successively slipped toward the east at speeds of about 30$-$40 km
s$^{-1}$ and seemed to peel off the flux rope. The slippage
exhibited a quasi-periodic pattern and the associated period was
about 130$\pm$30 s. The \emph{IRIS} slit crossed the west legs of
slipping structures and the spectroscopic observations of the Si
{\sc iv} 1402.77 {\AA} line showed a redshift of about 10$-$30 km
s$^{-1}$ and a line width reaching 50$-$120 km s$^{-1}$. The
quasi-periodic slipping motion lasted about 40 min and at the late
stage the flux rope started to rise up slowly. Then an X2.1 flare
occurred and almost simultaneously, the flux rope was impulsively
accelerated. One negative-polarity flare ribbon swept across a small
sunspot and the penumbral segments of the sunspot showed an evident
decay, with the \emph{IRIS} 2832 {\AA} emission intensity enhancing
by about 40 \% in about 50 min. The NLFFF extrapolation results show
the existence of a twisted flux rope at the flaring region. The
calculated 3D QSL structures are surrounding the flux rope,
indicating that the preflare slipping motion is around the
flux-rope-related QSLs.

The results suggest that quasi-periodic slipping magnetic
reconnection has started in the preflare stage. The Si {\sc iv}
profile was redshifted by 10$-$30 km s$^{-1}$ at the legs of
slipping structures, probably indicative of reconnection downflows.
The topology analysis of the 3D coronal magnetic field structure
shows that 3D QSL structures bind the twisted flux rope, separating
the flux rope from the surrounding field. The reconnection between
the field lines of the flux rope and neighboring field lines
continuously occurs along the flux-rope-related QSLs, causing the
exchange of field line linkage and the apparent slipping motion of
the flux rope structures. The quasi-periodic pattern indicates that
the slipping reconnection reappears for multiple times at a certain
location, with a regular time interval. The period of the preflare
slipping motion varies between 100 and 150 s, averaging $\sim$ 130
s, shorter than that of the slipping flare loops as reported by Li
\& Zhang (2015) and Li et al. (2015). The quasi-periodic
oscillations of the footpoints of slipping flare loops have been
reported by Brannon et al. (2015) and Brosius \& Daw (2015), who
analyzed the same event on 2014 April 18 using the \emph{IRIS} data.
Brannon et al. (2015) presented that the coherent oscillations of
small-scale substructure of the flare ribbon have an average period
of about 140 s, consistent with our observations. They interpreted
that a tearing mode or Kelvin-Helmholtz instability in coronal
current sheets drives these oscillations. Another explanation
responsible for oscillatory reconnection is related to the
relationship between energy load and unload balance (Nakariakov \&
Melnikov 2009; Brosius \& Daw 2015). In these models the magnetic
energy is continuously built up through photospheric (shearing and
converging) motions until a critical level is achieved resulting in
a release of the magnetic energy via reconnection. This process
repeats several times with the magnetic energy building up before
release.

The flux rope eruption and the X2.1 flare occurred just after the
slipping motion of flux rope structures. The long-duration slipping
magnetic reconnection at the border of the flux rope probably caused
the loss of equilibrium of the flux rope system, and drove the later
eruption of the flux rope and the associated flare. Magnetic
reconnection at the top border of the pre-eruptive flux rope has
rarely been reported. During the eruption process, most models
locate the reconnection site in the vertical current sheet below the
flux rope. Few observations showed that energy release and
reconnection could also occur at the leading edge of an erupting
flux rope (Ji et al. 2003; Huang et al. 2011; Jiang et al. 2016).
Wang et al. (2009) suggested that EUV brightenings at the far
endpoints of erupting filaments were caused by magnetic reconnection
at the boundary between the erupting filament and the background
corona. The helical current sheet at the interface between a
kink-unstable flux rope and the surrounding medium was simulated by
Kliem et al. (2004). The reconnection at the border of the flux rope
is different from the breakout model proposed by Antiochos et al.
(1999) in a closed quadrupolar configuration, although reconnection
sites are both overlying the erupting flux rope in the two
scenarios. The breakout model involves open fields for the
larger-scale connectivity region and the reconnection site is high
in the corona. However, the reconnection at the border of the flux
rope occurs at a lower altitude.

Our observations imply that the preflare slipping motion of flux
rope structures makes the flux rope system unstable, resulting in an
eruption. Preflare activities have been considered as a potential
clue to understand the triggering mechanism of solar flares, and
help to predict the occurrence of solar flares. Previous studies
showed that localized brightenings in X-ray, EUV/UV wavelengths and
microwave bursts occurred several minutes or more before the onset
of the flare (F{\'a}rn{\'{\i}}k et al. 1996; Chifor et al. 2007;
Zhang et al. 2015b). These preflare brightenings coincided with the
presence of emerging or cancelling flux (Joshi et al. 2011). Awasthi
et al. (2014) revealed the brightened loop top in 131 {\AA}
observations prior to the onset of the precursor phase. Our
observations also showed that the preflare brightenings at the loop
tops in 1400 {\AA} and 1330 {\AA} channels were more evident. Kim et
al. (2001) found a rapid flipping and connectivity change of
filament threads in H$\alpha$ images preceding the filament
eruption. The magnetic flipping might be associated with the
slipping magnetic reconnection in this work. It has been suggested
that the pre-flare activities may be a result of slow reconnection
and induce the filament eruption and the subsequent flaring (Moore
\& Roumeliotis 1992; Kim et al. 2001).

The observations of rapid penumbral decay during flares have been
reported before (Wang et al. 2004; Deng et al. 2005). It is
suggested that flare-related photospheric magnetic field changes
make the penumbral magnetic field relaxing upward by rapid magnetic
reconnection and becoming more vertical (Sudol \& Harvey 2005; Verma
\& Denker 2012). In our observations, the inflow towards the umbra
associated with the penumbral decay was first reported. When the
magnetic field lines in the penumbrae turned from more inclined to
more vertical, the material of the penumbral filaments flowed
towards their footpoints and resulted in the observed inflow towards
the umbra. Moreover, the area of the umbra decreased just after the
flare and then gradually increased during long-term evolution. The
area reduction of the umbra might be caused by the magnetic field
topology change during the flare. The later increase in the area of
the umbra was probably related to penumbral decay. For the magnetic
field lines in the penumbrae became vertical, the penumbrae at the
inner edge developed into the umbra that are mainly vertical
magnetic flux tubes.

\acknowledgments {\emph{SDO} is a mission of NASAs Living With a
Star Program. \emph{IRIS} is a NASA small explorer mission developed
and operated by LMSAL with mission operations executed at NASA Ames
Research center and major contributions to downlink communications
funded by the Norwegian Space Center (NSC, Norway) through an ESA
PRODEX contract. This work is supported by the National Natural
Science Foundations of China (11303050 and 11533008) and the
Strategic Priority Research Program$-$The Emergence of Cosmological
Structures of the Chinese Academy of Sciences, Grant No.
XDB09000000. K. Yang was supported by NKBRSF under grant
2014CB744203.}

{}
\clearpage

\begin{figure}
\centering
\includegraphics
[bb=45 125 507 689,clip,angle=0,scale=0.9]{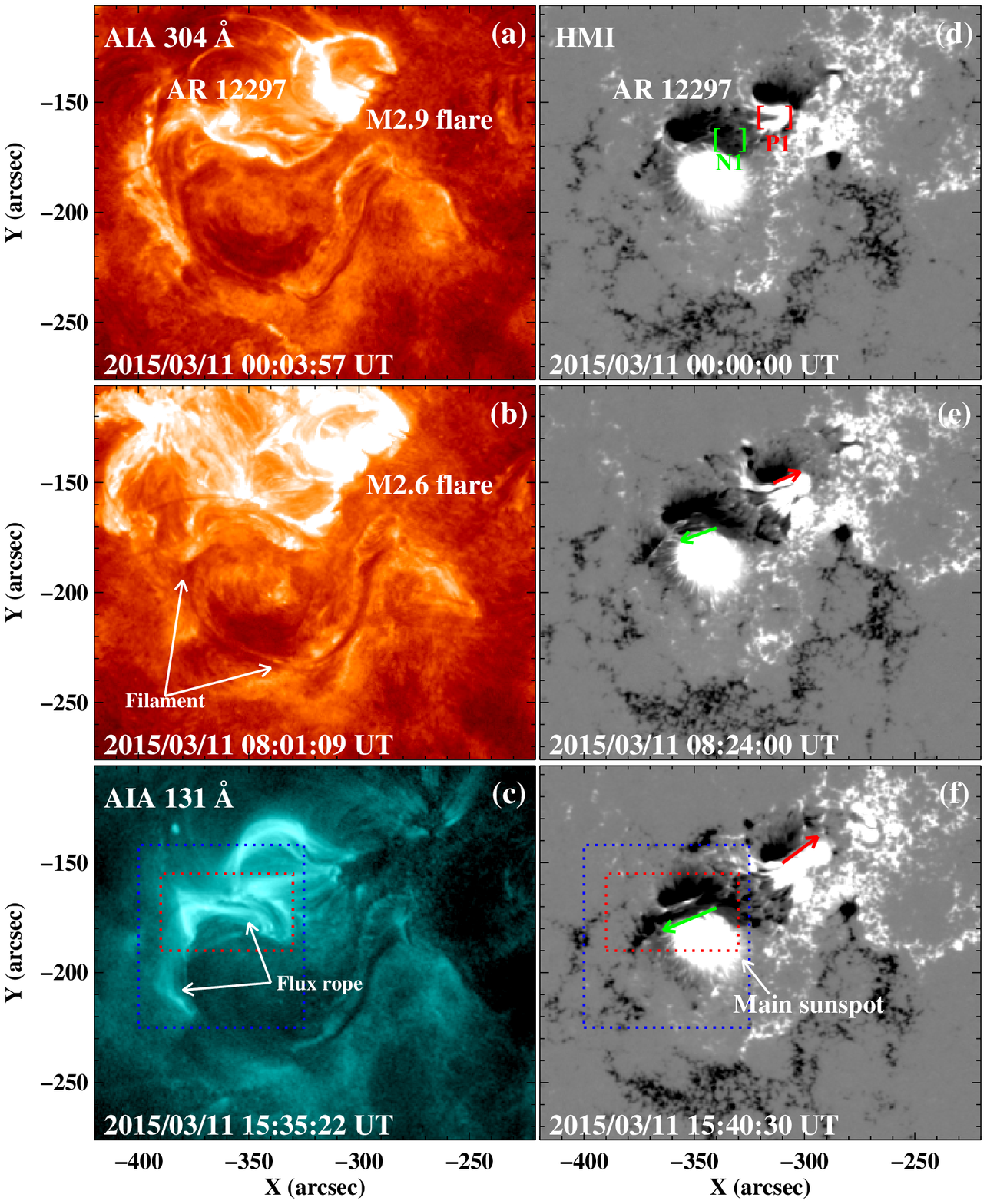} \caption{Time
sequence of AIA 304 {\AA} images, 131 {\AA} image and HMI LOS
magnetograms showing the evolving AR 12297 on 2015 March 11.
Positive patch P1 and negative patch N1 in panel (d) are an emerging
bipolar pair in the AR center. Red and green arrows respectively
denote the shearing motions of the emerging flux P1 and N1. The red
rectangles in panels (c) and (f) show the FOV of Figures 2$-$3, 5(a)
and 5(d). The blue rectangles show the FOV of Figure 7.
\label{fig1}}
\end{figure}
\clearpage

\begin{figure}
\centering
\includegraphics
[bb=46 144 511 670,clip,angle=0,scale=0.9]{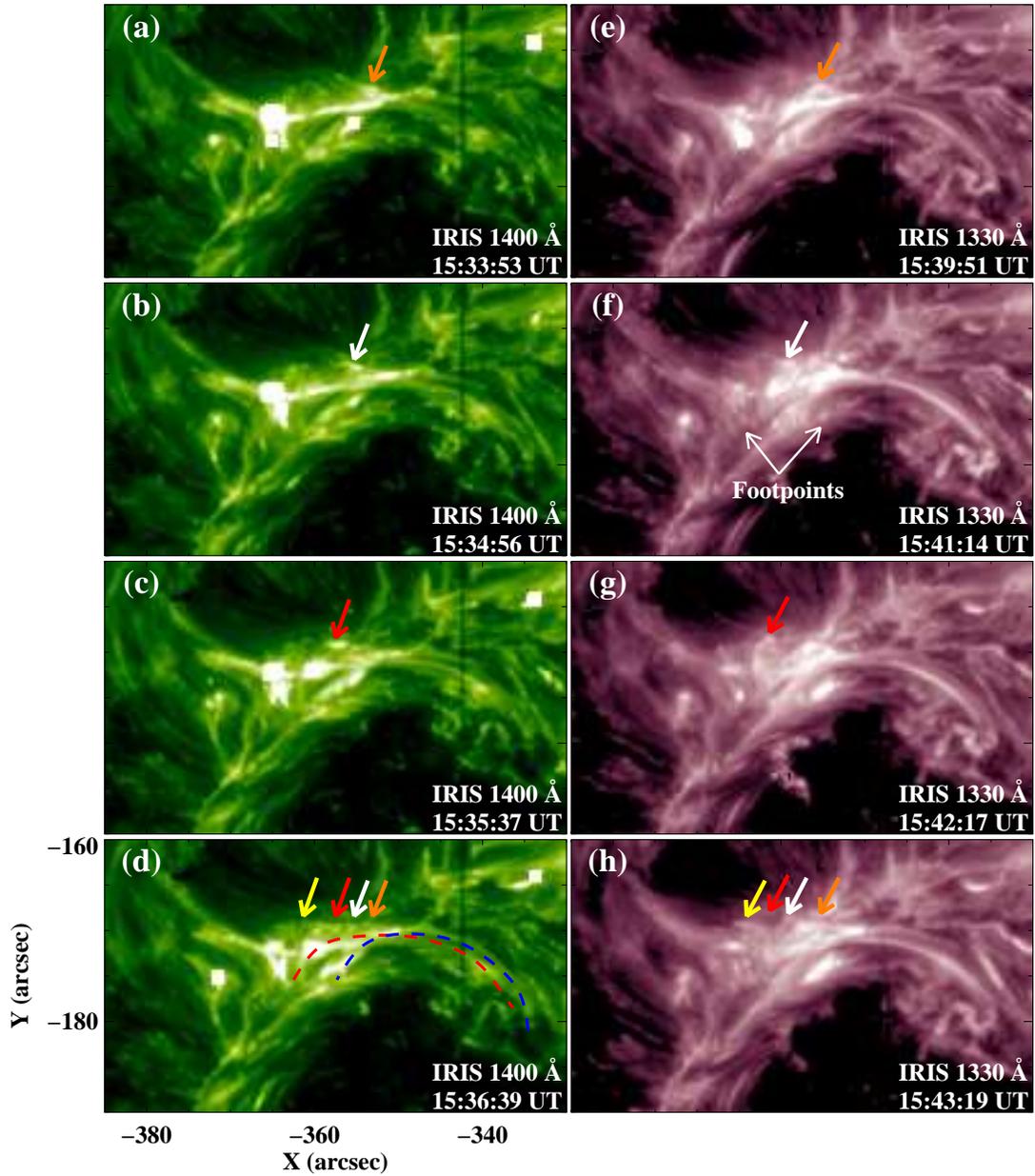}
\caption{Slipping motion of flux rope structures viewed in
\emph{IRIS} 1400 {\AA} and 1330 {\AA} images (see Animation
1400-slippage). The orange, white, red and yellow arrows in panels
(a)-(d) trace the apparent slipping motion of flux rope structures
from the west to the east. The west three arrows in panel (d) are
the duplicates of the arrows in panels (a)-(c). The arrows in panels
(e)-(h) represent another process of the slippage from the west to
the east and the meanings are the same as the left column. Red and
blue dashed curves in panel (d) outline two flux rope structures.
\label{fig2}}
\end{figure}
\clearpage

\begin{figure}
\centering
\includegraphics
[bb=46 206 510 610,clip,angle=0,scale=0.9]{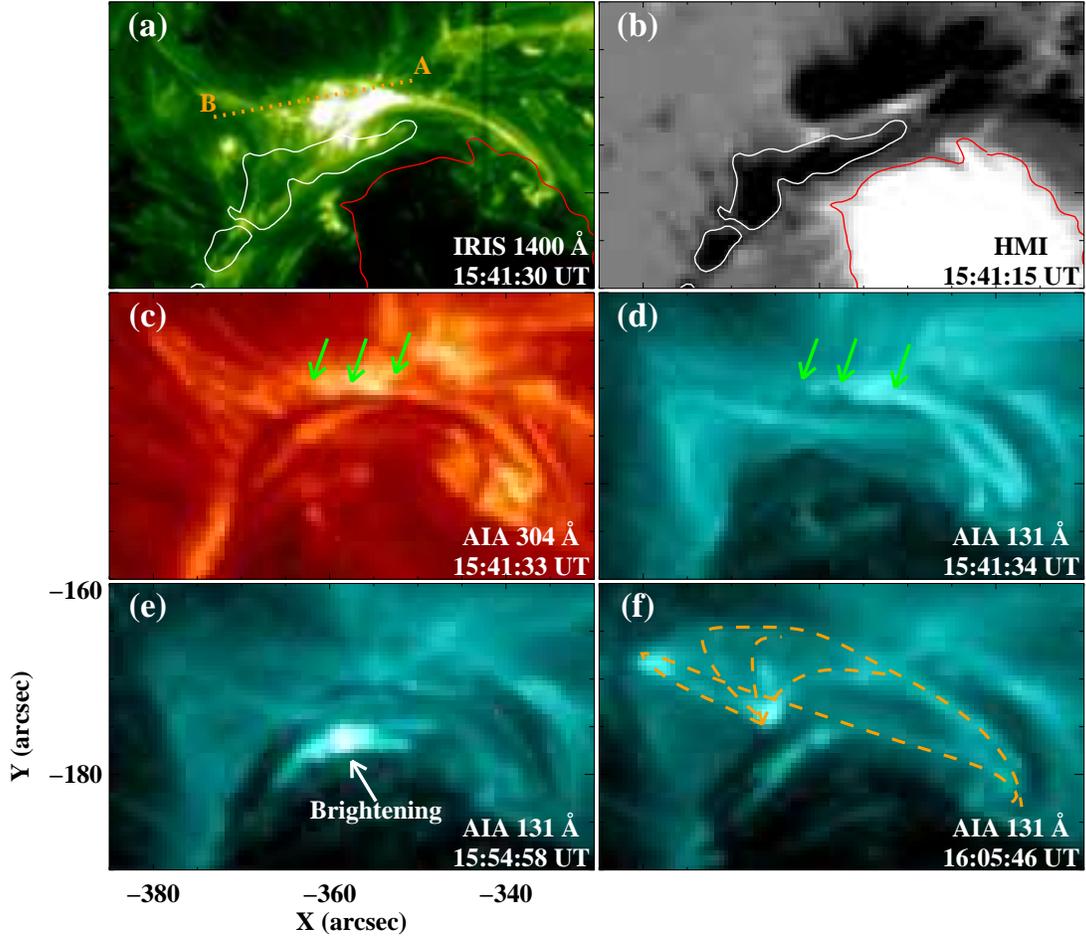}
\caption{Multi-wavelength observations and LOS magnetogram from the
\emph{IRIS} and \emph{SDO}. The red and white contours in panels
(a)-(b) are the magnetic fields at $\pm$ 500 G levels at the ends of
slipping structures. Dotted line ``A$-$B" (panel (a)) shows the cut
position used to obtain the time-distance plots shown in Figures
4(a)-(c). Green arrows in panels (c)-(d) point to the slipping dark
structures viewed in 304 {\AA} and 131 {\AA} images. The dashed
curves in panel (f) delineate a fan-shaped surface at the late stage
of slipping motion. \label{fig3}}
\end{figure}
\clearpage

\begin{figure}
\centering
\includegraphics
[bb=54 207 507 682,clip,angle=0,scale=0.9]{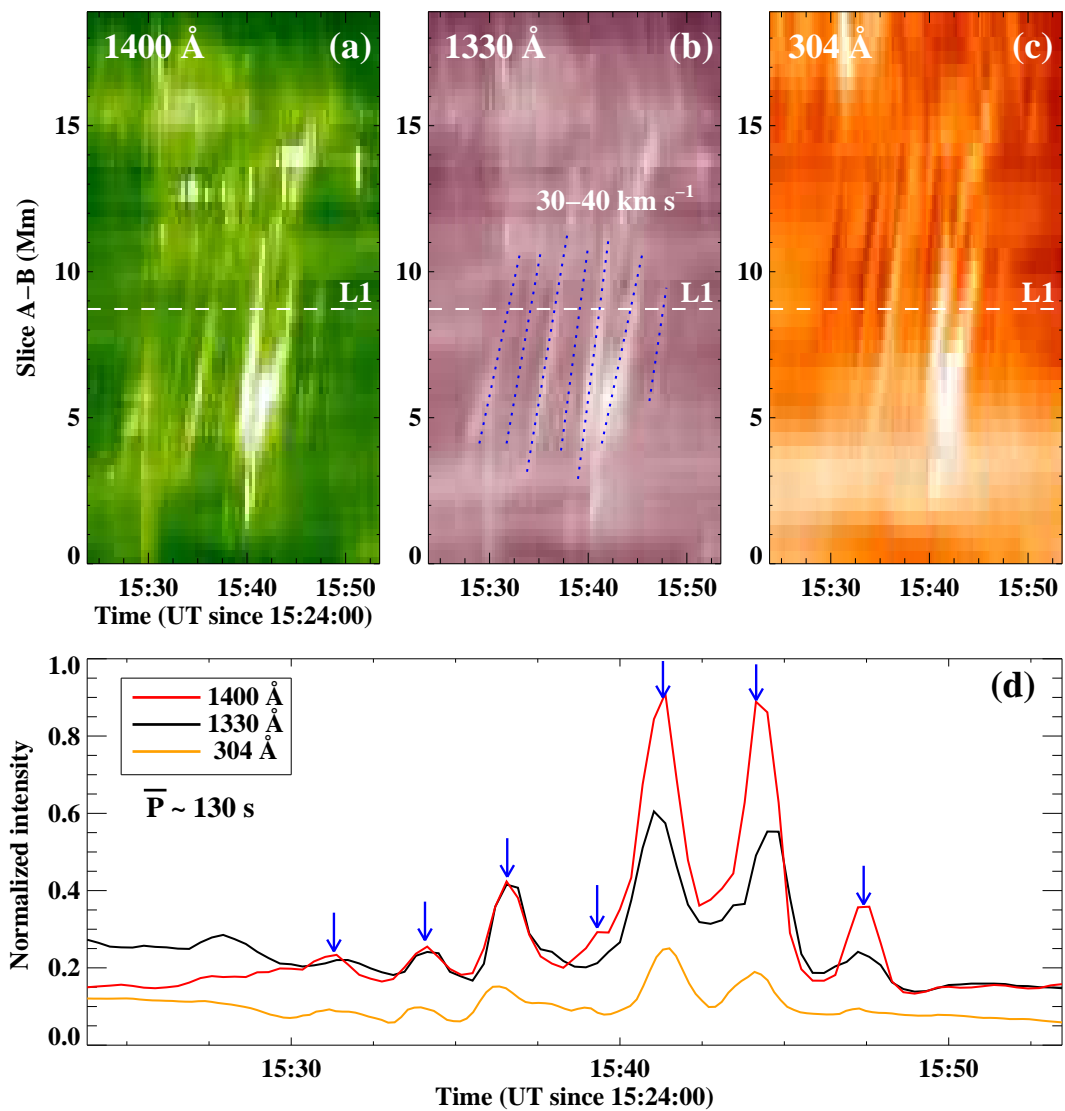} \caption{Panels
(a)-(c): time-distance plots along slice ``A$-$B" (dotted line in
Figure 3(a)) at 1400 {\AA}, 1330 {\AA} and 304 {\AA} showing the
quasi-periodic slippage of flux rope structures. Panel (d):
horizontal slices along the dashed lines (``L1") in panels (a)-(c).
$\overline{P}$ represents the average period ($\sim$ 130 s), and
blue arrows point to the peaks of intensity at 1400 {\AA}.
\label{fig4}}
\end{figure}
\clearpage

\begin{figure}
\centering
\includegraphics
[bb=14 296 489 524,clip,angle=0,scale=0.9]{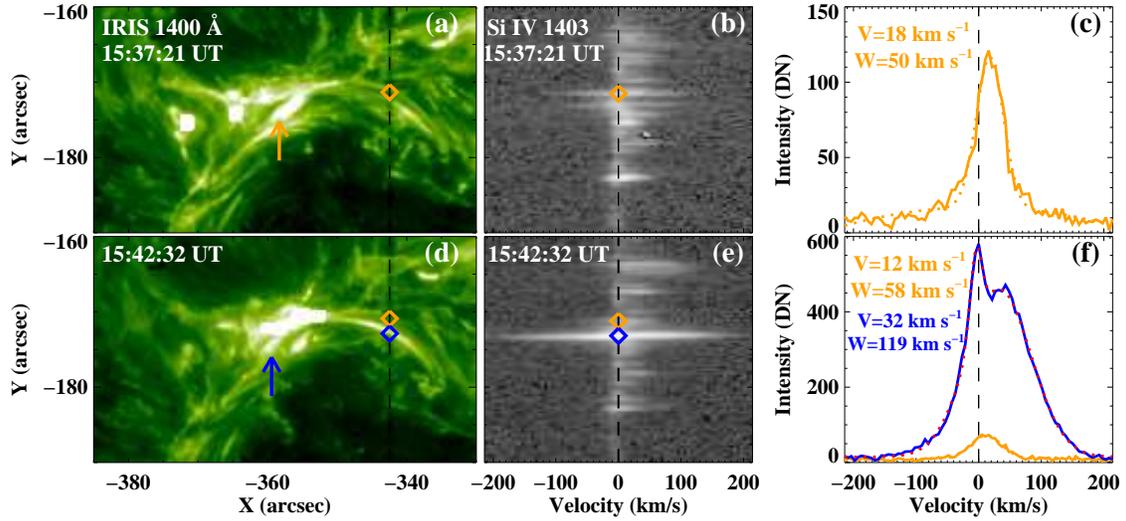} \caption{Left
column: \emph{IRIS} 1400 {\AA} images displaying the slipping flux
rope structures. The orange and blue diamonds denote the
intersections of the loop-like structures and the slit. The arrows
point to the brightenings at the east footpoints of analyzed
structures. Middle column: simultaneous appearance of the Si {\sc
iv} 1402.77 {\AA} spectra in the slit range of panels (a) and (d).
Right column: profiles of the Si {\sc iv} line at the selected
locations. The orange dotted curves are the single-Gaussian fitting
profiles. The red dotted curve is the double-Gaussian fitting
profile. \label{fig5}}
\end{figure}
\clearpage

\begin{figure}
\centering
\includegraphics
[bb=11 1 492 300,clip,angle=0,scale=0.9]{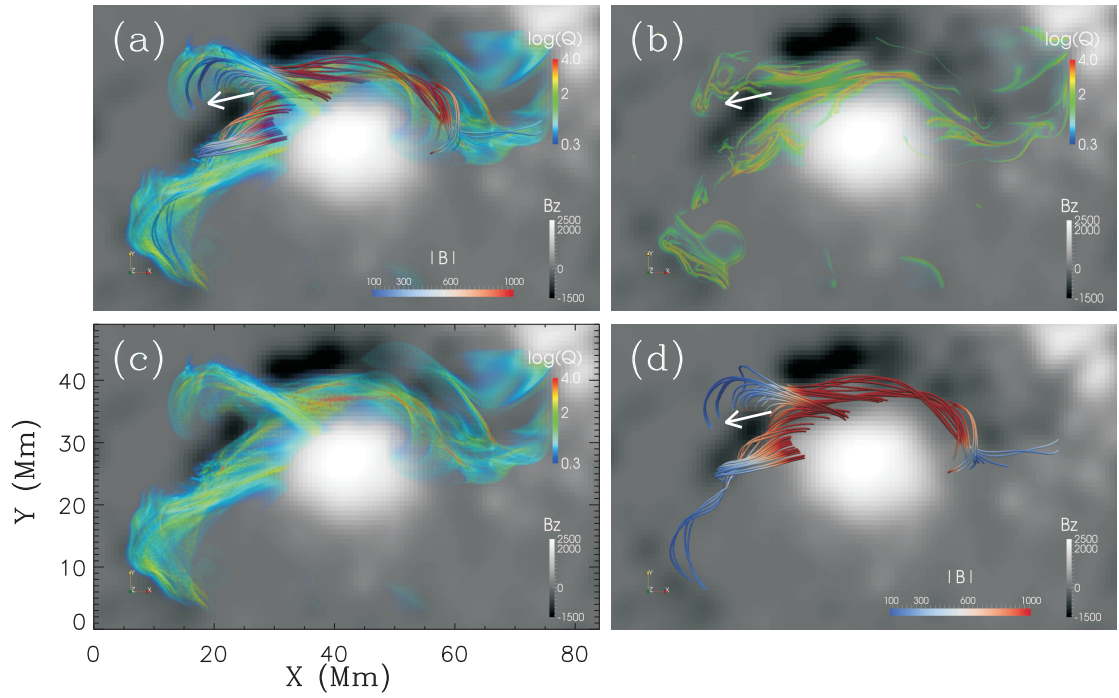} \caption{Topology
of the extrapolated magnetic field. Panels (a), (c), and (d): field
lines of the flux rope and its associated 3D QSL. Panel (b): The
intersection of the QSL with the bottom boundary. The background
shows the photospheric vertical magnetic field at 16:04 UT. White
arrows represent the slipping motion of flux rope structures before
the flare. \label{fig8}}
\end{figure}
\clearpage

\begin{figure}
\centering
\includegraphics
[bb=41 161 517 652,clip,angle=0,scale=0.9]{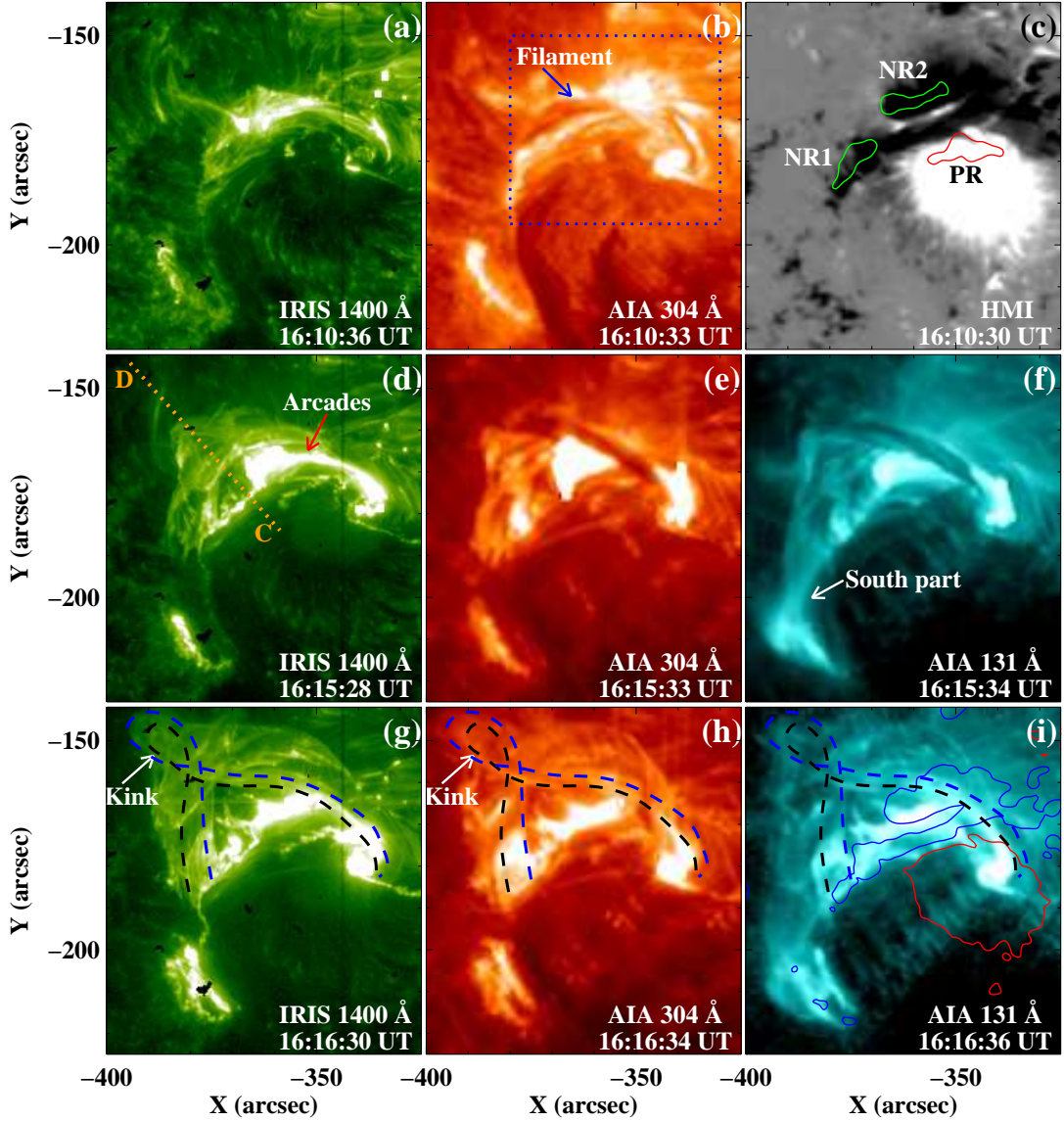}
\caption{Eruption of the flux rope observed by the \emph{IRIS} and
\emph{SDO} (see Animations 1400-eruption and 131-eruption). The blue
rectangle in panel (b) denotes the FOV of Figures 9(a)-(f). The red
and green curves in panel (c) are the contours of brightenings in
the 1600 {\AA} image at the flare peak time, displaying one positive
(PR) and two negative flare ribbons (NR1 and NR2). Dotted line
``C$-$D" (panel (d)) shows the cut position used to obtain the
time-distance plots shown in Figure 8. Blue and black curves in
panels (g)-(i) outline partial structures of the flux rope. The red
and blue contours in panel (i) are the magnetic fields at $\pm$ 450
G levels. \label{fig7}}
\end{figure}
\clearpage

\begin{figure}
\centering
\includegraphics
[bb=96 144 505 678,clip,angle=0,scale=0.9]{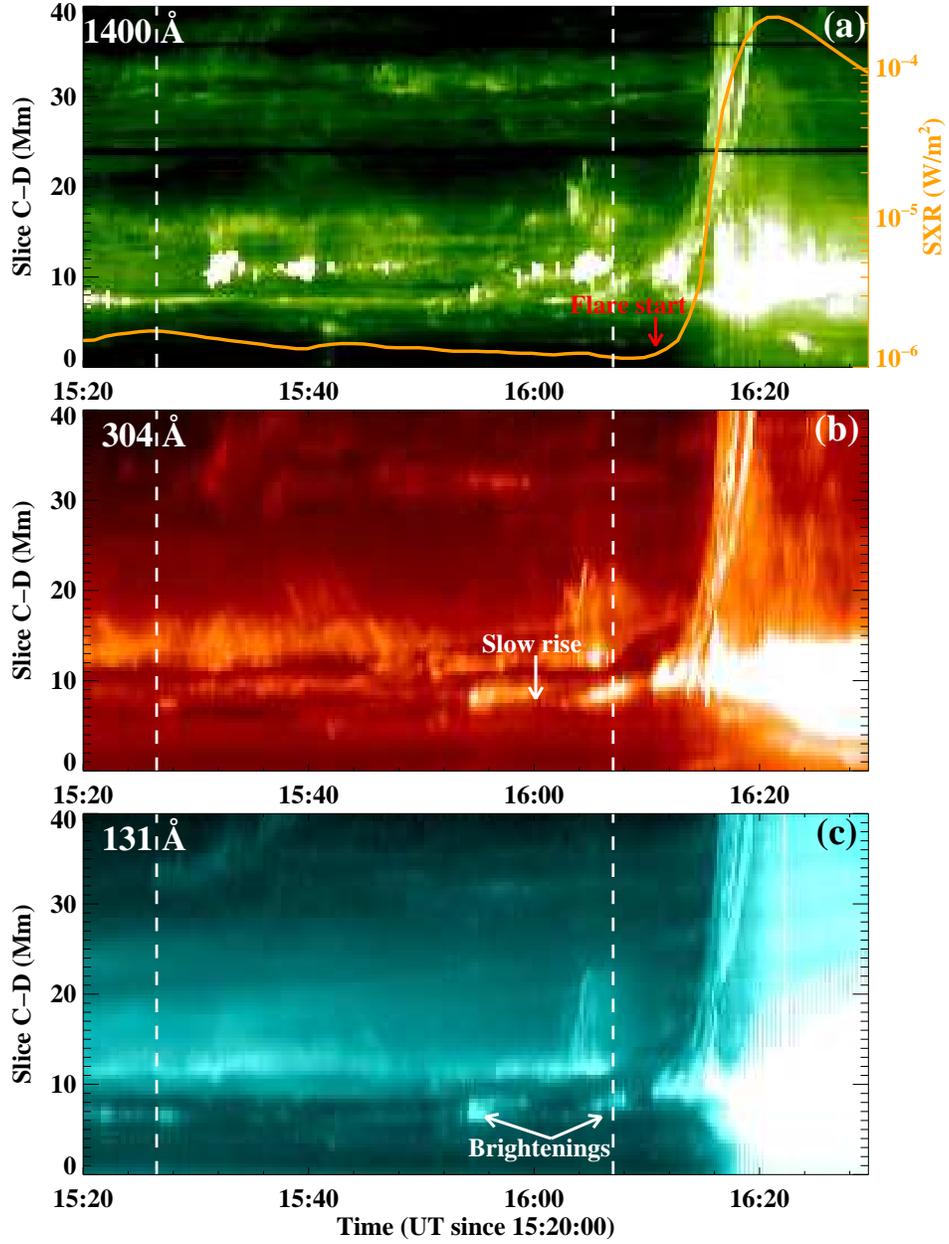}
\caption{Time-distance plots along slice ``C$-$D" (dotted line in
Figure 7(d)) at 1400 {\AA}, 304 {\AA} and 131 {\AA} showing the
kinematic evolution of the erupting flux rope. The orange curve in
panel (a) shows GOES SXR 1-8 {\AA} flux of the associated X2.1
flare. Two dashed lines in each panel denote the time interval
between 15:27 UT and 16:07 UT when the slipping motion of flux rope
structures occurred. \label{fig8}}
\end{figure}
\clearpage

\begin{figure}
\centering
\includegraphics
[bb=24 156 536 636,clip,angle=0,scale=0.9]{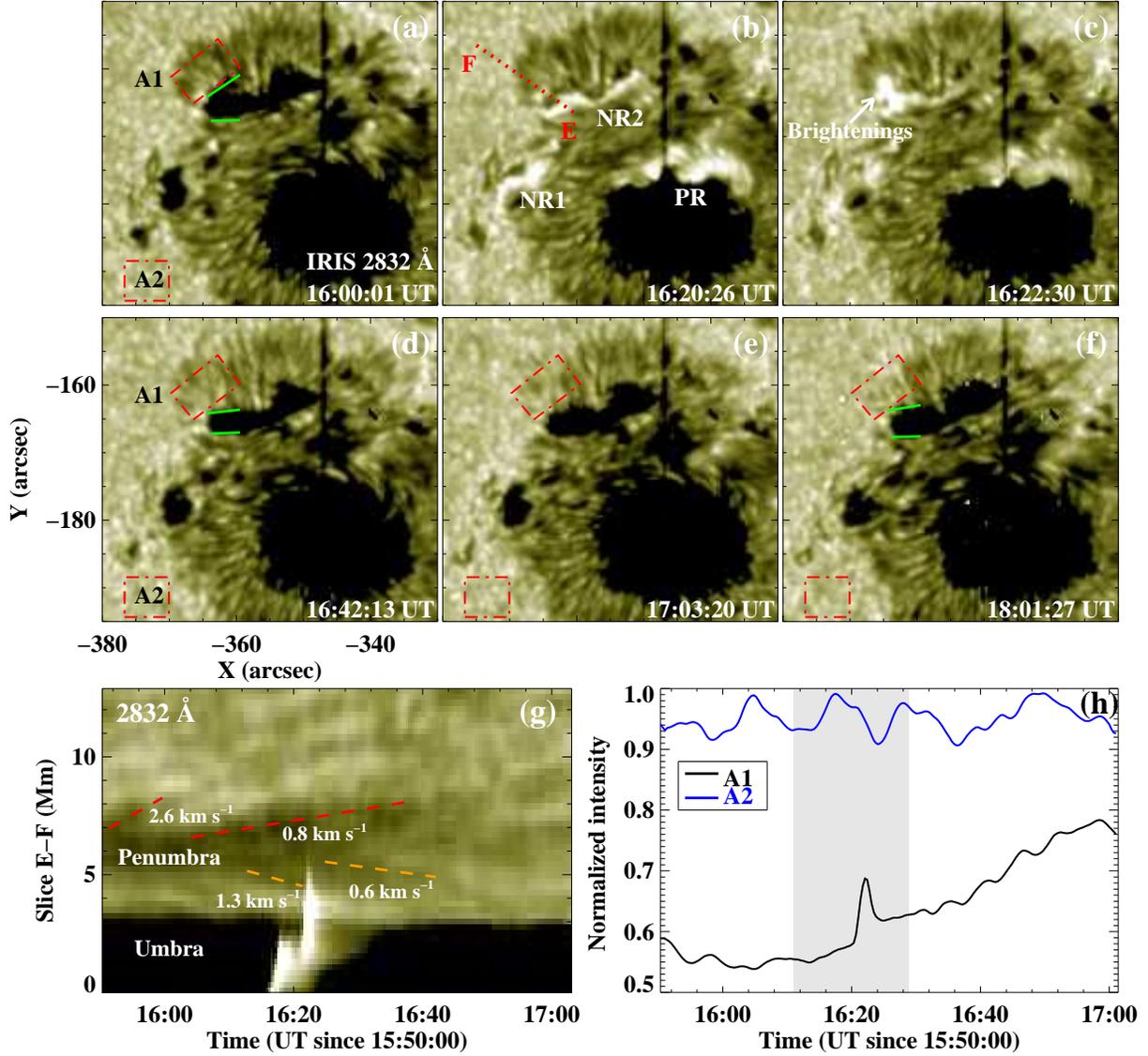} \caption{Panels
(a)-(f): \emph{IRIS} 2832 {\AA} images showing the penumbral decay
following the X2.1 flare. Areas ``A1" and ``A2" denote the penumbra
and background regions where the intensity-time profiles in panel
(h) are calculated. Green lines represent the size of the umbra
before and after the flare. Dotted line ``E$-$F" (panel (b)) shows
the cut position used to obtain the time-distance plot shown in
panel (g). Panels (g)-(h): time-distance plot along slice ``E$-$F"
at 2832 {\AA} and intensity-time profiles within regions ``A1" and
``A2". Red dashed lines in panel (g) indicate the penumbra outflow
and orange dashed ones indicate the inflow towards the umbra after
the flare start. Gray section in panel (h) represents the time
interval between the flare start and end. \label{fig9}}
\end{figure}
\clearpage

\end{document}